\documentclass{jetpl}
\twocolumn

\lat

\title{Scroll-wave dynamics in the presence of ionic and conduction
inhomogeneities in an anatomically realistic mathematical model for the pig
heart}

\rtitle{Scroll-wave dynamics in an anatomically\ldots}

\sodtitle{Scroll-wave dynamics in the presence of ionic and conduction inhomogeneities in an anatomically realistic mathematical model for the pig heart}

\author{R.\, Majumder$^a$, Rahul\, Pandit$^{b}$, A.\,V.\, Panfilov$^{c,*}$}

\rauthor{R.\, Majumder, Rahul\, Pandit, A.\,V.\, Panfilov}

\sodauthor{Majumder, Pandit, Panfilov}

\address{${^a}$Laboratory of Experimental Cardiology, Dept. of Cardiololgy, Heart Lung Center, Leiden University Medical center,
2333ZA Leiden, the Netherlands\\~\\
${^{b}}$Centre for Condensed Matter Theory, Dept. of Physics, Indian Institute of Science, Bangalore 560012, India; Jawaharlal Nehru Centre for Advanced Scientific Research, Bangalore, India.\\~\\
${^c}$Dept. of Physics and Astronomy, Gent University, Krijgslaan 281, S9, 9000 Gent, Belgium;
Moscow Institute of Physics and Technology, (State University), Dolgoprudny, Moscow Region, Russia\\
$^*$e-mail: Alexander.Panfilov@ugent.be}

\dates{22 August 2016}

\abstract{Nonlinear waves of the reaction-diffusion (RD) type occur in many 
biophysical systems, including the heart, where they initiate cardiac 
contraction. Such waves can form vortices called scroll waves, which result 
in the onset of life-threatening cardiac arrhythmias. The dynamics of 
scroll waves is affected by the presence of inhomogeneities, which, in a 
very general way, can be of \textit{(i)} ionic type, i.e., they affect 
the reaction part, or \textit{(ii)} conduction type, i.e., they affect 
the diffusion part of an RD equation. We demostrate, for the first 
time, by using a state-of-the-art, anatomically realistic model of the 
pig heart, how differences in the geometrical and biophysical nature of 
such inhomogeneities can influence scroll-wave dynamics in different ways. 
Our study reveals that conduction-type inhomogeneities become increasingly 
important at small length scales, i.e., in the case of multiple, randomly 
distributed, obstacles in space at the cellular scale ($0.2-0.4{\rm mm}$). 
Such configurations can lead to scroll-wave break up. In contrast,
ionic inhomogeneities, affect scroll-wave dynamics significantly
at large length scales, when these inhomogeneities are 
localized in space at the tissue level ($5-10$ mm). In such configurations, 
these inhomogeneities can (a) attract scroll waves, by pinning them to the 
heterogeneity, or (b) lead to scroll-wave breakup.}


\begin{document}

\maketitle

\textbf{Introduction:}
Nonlinear waves occur in excitable media of physical, chemical, and 
biological origin. Such waves can form vortices in two and three dimensions;
these are called spiral and scroll waves, respectively, and they  
are involved in the spatiotemporal organization of wave dynamics in various 
complex systems. Therefore, the study of such waves is a subject of interest 
in a broad area of research. One of the most important applications of such
studies is the formation of vortices in cardiac tissue, which is associated 
with the onset and development of lethal cardiac 
arrhythmias~\cite{clayton10,tks07,tks09,Jalife98,Mann02,Kleber04,Vigmond08}. 
Thus, understanding the factors that determine the dynamics of scroll waves 
is a topic of great interest. Cardiac arrhythmias, such as ventricular 
tachycardias (VT) are generally associated with stationary, meandering, or 
drifting, periodic or quasiperiodic scroll waves; whereas, ventricular 
fibrillation (VF) is associated with scroll-wave break up. The dynamical 
behaviour of scroll waves in cardiac tissue is affected significantly
by the presence of inhomogeneities
~\cite{Ikeda97,Valderrabano00,Lim06,Davidenko92,tks07,tks09,Majumder11,Majumder12,Nayak13,
Rudenko83,Panfilov91}, which can occur in the heart in many forms. However, 
biophysically, they can be grouped into two major classes: \textit{(i)} 
Ionic-type, i.e., inhomogeneities in the properties of different cells that 
constitute the system; and \textit{(ii)} conduction-type, i.e., 
inexcitable obstacles. An in-depth knowledge of the role of these 
inhomogeneities is essential for understanding the mechanisms that underlie 
most cardiac arrhythmias. 

In experiments, it is often difficult to study systematically the role of 
these inhomogeneities in the development of arrhythmias, with regard to the 
nature, position, and distribution of these inhomogeneities within the heart. 
Thus, it is important to search for alternative methods of investigation. 
Mathematical modelling provides an important tool here; it has been used 
extensively, with outstanding success, in interdisciplinary 
science. From a mathematical point of view, the excitable, cardiac-tissue 
medium is described by a reaction-diffusion (RD) equation of the type:\\
\begin{equation}
  \frac{\partial v}{\partial t} = \nabla . (\mathcal{D}\nabla v) + \mathcal{F}(g,v),\\
\end{equation}
with the reaction part $\mathcal{F}(g,v)$ accounting for properties of 
cardiac cells and the diffusion part $\nabla . \mathcal{D}\nabla v$, for 
the connection of cells to tissue. In this setting, an ionic 
inhomogeneity represents a modification of $\mathcal{F}$, whereas a 
conduciton inhomogeneity involves a modification of 
$\nabla . \mathcal{D} \nabla v$. 

In this Letter, we present an extensive numerical study of scroll-wave dynamics
in the presence of inhomogeneities in an anatomically realistic model of the
pig heart. We have used the single-cell, modified, Luo-Rudy I (mLRI)
model~\cite{Qu99} to construct our cardiac-tissue model and the anatomically
realistic geometry obtained in~\cite{Stevens03}.  We have studied the effects
on scroll-wave dynamics of \textit{(i)} large-length-scale, solitary
inhomogeneity (old infarction) and \textit{(ii)} small-length-scale, multiple,
conduction inhomogeneities (fibrosis) and compared our results from these
studies with those we have obtained from similar \textit{(i)} large- and
\textit{(ii)} small-length-scale ionic inhomogeneities. Our results illustrate,
for the first time, that conduction inhomogeneities influence scroll-wave
dynamics significantly, when they occur at small length scales (sub-millimeter)
in distributed patterns; by contrast, ionic inhomogeneities play a significant
role in influencing such dynamics at large length scales (millimeters), when
they are localized in space.\\

\textbf{Methods:} 
A modified version of the original Luo-Rudy I model~\cite{LRI91}, namely, the
mLRI~\cite{Qu99}, was used to model the electrophysiological properties of the
pig cardiac cell. The original parameters of the mLRI model, including the
effects of Eqs.4-7 of Qu, \textit{et al.}~\cite{Qu99} were used to simulate the
pig heart electrophysiology in our studies. In two dimensions (2D), this
parameter set yielded a spiral wave rotating at a frequency $\simeq12{\rm Hz}$,
the approximate frequency of spiral waves~\cite{panfilov06,Newton04} in the pig
heart. 

Here the transmembrane potential ($V$) of a cardiac cell 
depends on the sum of $6$ ionic currents ($I_{ion}$) and the applied current
stimulus ($I_{stim}$) according to the following partial differential equation:
\begin{eqnarray}
\frac{\partial V}{\partial t} = \nabla \cdot (\mathcal{D}\nabla V) - \frac{I_{ion} + I_{stim}}{C} ,
\label{pde}
\end{eqnarray}
where $C$ is the specific membrane capacitance of the cell. The diffusion
tensor $\mathcal{D}$ is a $3\times 3$ matrix~\cite{Clayton08,tenTusscher07}
with elements
\begin{eqnarray}
\mathcal{D}_{ij} = D_{\parallel} * \delta_{ij} + 
(D_{\parallel} - D_{\perp})\alpha_i\alpha_j.
\label{diffusion tensor}
\end{eqnarray}
Diffusion coefficients for longitudinal ($D_{\parallel}$) and transverse
($D_{\perp}$) propagation are chosen as $0.001 cm^2/ms$ $0.00025 cm^2/ms$,
respectively, to obtain conduction velocities $\simeq 59 cm/s$ and $\simeq 20.5
cm/s$, respectively, in the longitudinal and transverse directions; these are
consistent with the normally accepted values for pig cardiac
tissue~\cite{Kleber86}. The vector $\alpha$ specifies the local, muscle-fiber
orientation.  

To construct an anatomically realistic simulation domain, processed DTMRI data
points, have been embedded into a cubical simulation
domain~\cite{Clayton08,tenTusscher07}, with $328^3$ vertices. Each node in this
cubical domain are labeled as a heart point (HP), if the node coincides with
one of the points from the processed data set, or as a non-heart point (NHP)
otherwise. The temporal part of Eq.~\ref{pde} is solved by using Euler's
method; we use a centered, finite-difference scheme with $\delta x = \delta y =
\delta z = 0.025$~cm to solve Eq.~\ref{pde} in space. Zero-flux boundary
conditions are incorporated on the boundaries of the anatomically realistic
heart by adopting a phase-field approach~\cite{Fenton05}. 

\begin{figure}[!htbp]
\includegraphics[width=82mm]{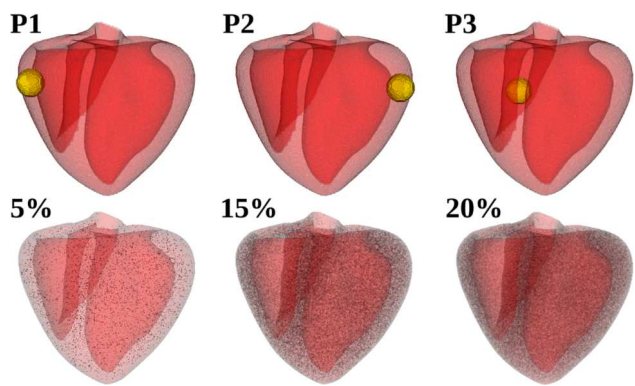}
\caption{Figure~\ref{inhoms}: The upper panel shows schematic diagrams of the
pig heart with the large-length-scale, spherical inhomogeneity at positions
\textbf{P1}, \textbf{\bf P2}, and \textbf {P3}.  The lower panel shows
small-length-scale inhomogeneities that are distributed randomly, covering
$5\%, \; \; 15\%$, and $20\%$, respectively, of nodal sites.}
\label{inhoms}  
\end{figure}

To model large-scale inhomogeneities, spheres of radius $10\delta x$ were
embedded in $3$ different positions of the simulated heart wall ($P1$, $P2$,
and $P3$), with the possibility of protrusion into the inner cavities, or out
of the exterior surface of the heart. Small-scale inhomogeneities were modeled
as randomly distributed cubical cells of side $1\delta x$~\cite{tenTusscher07},
that contained $1$ node each ( $5\%, \; 10\%,\; 15\%$, and $20\%$ by number).
To model conduction-type inhomogeneities, $D_\parallel=D_\perp$ was set to $0$
inside the inhomogeneity. To model ionic inhomogeneities, only the value of the
slow, inward conductance $G_{si}$ was set to $0.02{\rm mS/cm^2}$ ~\cite{tks09}
at the sites covered by the inhomogeneity, without adjusting the elements of
the diffusion tensor.  Figure~\ref{inhoms} shows the positions and
distributions of inhomogeneities considered.\\

\begin{figure}[!htbp]
\includegraphics[width=82mm]{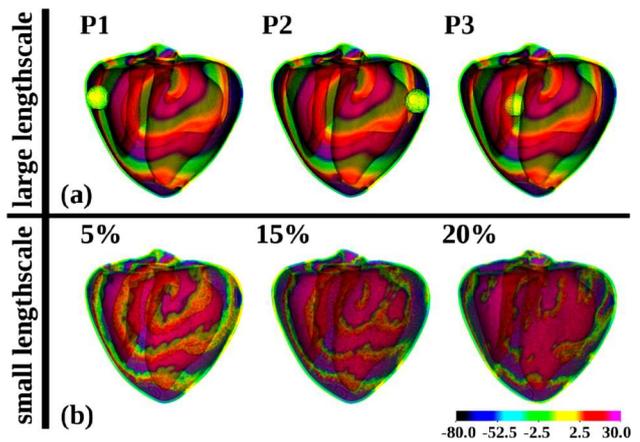}
\caption{Figure~\ref{cond_inhom}: Representative snapshots of 
scroll-wave dynamics in anatomically 
realistic pig hearts in the presence of (a) large- and 
(b) small-scale (b), conduction inhomogeneities.}
\label{cond_inhom}  
\end{figure}

\textbf{Results:} \textbf{\textit{No inhomogeneities:}} In the absence of inhomogeneities, we obtain a single stable periodically
rotating scroll, with an average frequency $12{\rm Hz}$.  We then study
scroll-wave dynamics in the presence of large- and small-scale conduction and
ionic inhomogeneities.  Thus, in total, we consider $10$ different cases with
different inhomogeneities. Our main findings from these $10$ cases are listed
in Tables I \& II. The details of our results are also discussed below with
figures to illustrate the most important types of dynamical behaviours.\\
\\
\textbf{\textit{Conduction Inhomogeneities:}} Figure~\ref{cond_inhom} shows the effects of various conduction heterogeneities
on scroll-wave dynamics (cases 1-2). We find that solitary, large-scale
conduction inhomogeneities do not have any pronounced effect on scroll-wave
dynamics. Indeed, at all $3$ positions of the inhomogeneity $P1, P2$ and $P3$,
the scroll wave remains insensitive to the presence of the obstacle
(Figures~\ref{cond_inhom} (a)). However, small-scale conduction inhomogeneities
affect scroll-wave dynamics substantially by changing the characteristics of
the scroll wave and causing its breakup.  At all distributions of small-scale
conduction inhomogeneities that we have considered, namely, $5\%,10\%,15\%$ and
$20\%$ inhomogeneity (Case 2), we observe the following: \textit{(i)} a
shortening of the spatial wavelength of the scroll, and \textit{(ii)}
scroll-wave breakup at inhomogeneities $\gtrsim$ 15\% (Figure~\ref{cond_inhom}
(b)).\\

\begin{tabular}{|p{0.6 cm}|p{2.2 cm}|p{4.3 cm}|}
 \hline
 \multicolumn{3}{|c|}{Table I: Conduction inhomogeneity} \\
 \hline
 Case no.& Inhomogeneity type & Dynamics\\
 \hline
 1 & large ($P1$) & Scroll wave remains passive\\
   & large ($P2$) & towards the presence of the\\
   & large ($P3$) & inhomogeneity.\\
 \hline
 2 & small ($5\%$) & Scroll wavelength reduces.\\
   & small ($10\%$) & Scroll wavelength reduces.\\
   & small ($15\%$) & Unstable breakup.\\
   & small ($20\%$) & Stable breakup.\\   
 \hline
\end{tabular}

\begin{figure}[!htbp]
\includegraphics[width=82mm]{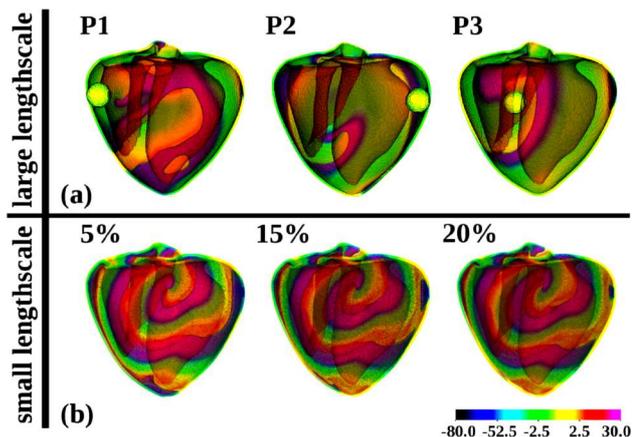}
\caption{Figure~\ref{ionic_inhom}: Representative snapshots of scroll-wave 
dynamics in anatomically realistic pig hearts in the presence of (a) large- 
and (b) small-scale (b) ionic inhomogeneities.} 
\label{ionic_inhom}  
\end{figure}

\textbf{\textit {Ionic Inhomogeneities:}} Figure~\ref{ionic_inhom} illustrates the effects of various ionic
heterogeneities on scroll-wave dynamics (Cases 3-4).  We see that solitary,
large-scale ionic inhomogeneities have a substantial effect on scroll-wave
dynamics.  We observe interesting dynamical behaviour, such as, scroll-wave
breakup (Figure~\ref{ionic_inhom}(a):\textbf{P1}) and anchoring \textbf{P3})
(Case 3).  On the contrary, small-scale ionic inhomogeneities do not lead to
qualitatively interesting dynamics: for all the inhomogeneities we have
considered, namely, $5\%,10\%,15\%$ and $20\%$ (Case 4)
(Figure~\ref{ionic_inhom} (b)) we do not find a pronounced change in
scroll-wave dynamics.\\

\begin{tabular}{|p{0.6 cm}|p{2.2 cm}|p{4.3 cm}|}
 \hline
 \multicolumn{3}{|c|}{Table II: Ionic inhomogeneity} \\
 \hline
 Case no.& Inhomogeneity type & Dynamics\\
 \hline
 3 & large ($P1$) & Stable breakup.\\
   & large ($P2$) & No change.\\
   & large ($P3$) & Stable anchoring.\\
 \hline
 4 & small ($5\%$) & No significant qualitative\\
   & small ($10\%$) & change. Dynamics is \\
   & small ($15\%$) & insensitive to the presence \\
   & small ($20\%$) & of the inhomogeneity.\\
 \hline
\end{tabular}

Taken together, our results demonstrate that large-scale conduction
inhomogeneities do not affect scroll-wave dynamics in the pig heart. However,
if the inhomogeneity is of ionic type, it can lead to scroll-wave breakup.
On the contrary, small-scale conduction inhomogeneities have a significant
influence on the dynamics of scroll waves; such inhomogeneities generally
lead to some decrease in the spatial wavelength of the scroll wave and 
initiate scroll-wave breakup. Small-scale ionic inhomogeneities, however, 
prove to be protective against breakup.\\

\textbf{Discussion:}We have carried out a comprehensive numerical study 
of scroll-wave dynamics in an ionic model for pig cardiac tissue; and 
we have compared, in the same conditions, the effects of conduction 
and ionic heterogeneities, both for small and large length scales~\cite{tks09,Majumder11,Majumder12}, on such scroll-wave dynamics.\\
Our principal, qualitative result that small-scale inhomogeneities are 
important in the diffusion part is a consequence of the effect of the 
diffusion processes on the reaction part (called the electrotonic effect 
in electrophysiology)~\cite{Defaw13}. However, we have also found that 
small-scale conduction inhomogeneities are not averaged out by the diffusion.
Therefore, their mean-field consideration, e.g., by using homogenization 
techniques, should be done with caution. For large-scale heterogeneities, 
our results are in line with findings for human cardiac-tissue 
simulations~\cite{Defaw14}; however, these have been 
performed on a completely different cardiac geometry, different cell models, 
and for substantially different values of scroll wavelengths. In addition 
to dynamical anchoring (via the transient-breakup phase) described 
in Ref.~\cite{Defaw14} we have also observed anchoring of the other type 
resulting from a drift of the scroll for qualitatively different positions of 
the heterogeneity: in particular, we have placed heterogeneity inside the 
septum and have found that it can attract scroll waves 
and thus lead to interesting new dynamics. 
\\

The work of RM and RP was supported by DST, UGC, and CSIR (India). RM and RP would like to thank SERC (IISc) for computational resources.
The work of AVP was supported by the Research Foundation-Flanders (FWO Vlaanderen).

\end{document}